\documentclass[
    ,final            % use final for the camera ready runs
  ]
  {aipproc}

\layoutstyle{6x9}

\begin{document}

\title{The jet/disk connection in blazars}

%  http://www.aip.org/pacs/index.html

\author{Gabriele Ghisellini}{
  address={INAF -- Osservatorio Astronomico di Brera, 
Via Bianchi 46, I--23807 Merate, Italy}
}

\begin{abstract}

The new high energy data coming mainly from the {\it Fermi} and 
{\it Swift} satellites and from the ground based Cerenkov telescopes
are making possible to study not only the energetics of blazar jets,
but also their connection to the associated accretion disks.
Furthermore, the black hole mass of the most powerful objects
can be constrained through IR-optical emission, originating 
in the accretion disks. 
For the first time, we can evaluate jet and accretion powers in units
of the Eddington luminosity for a large number of blazars.
Firsts results are intriguing. 
Blazar jets have powers comparable to, and often larger than the 
luminosity produced by their accretion disk. 
Blazar jets are produced at all accretion rates (in Eddington units), 
and their appearance 
depends if the accretion regime is radiatively efficient or not.
The jet power is dominated by the bulk motion of matter, not  by
the Poynting flux, at least in the jet region where the bulk of the
emission is produced, at $\sim 10^3$ Schwarzschild radii.
The mechanism at the origin of relativistic jets must be 
very efficient, possibly more than accretion, even if
accretion must play a crucial role.
Black hole masses for the most powerful jets at 
redshift $\sim3$ exceed one billion solar masses,
flagging the existence of a very large population of heavy
black holes at these redshifts.
\end{abstract}

\maketitle

%%%%%%%%%%%%%%%%%%%%%%%%%%%%%%%%%%%%%%%%%%%%
%% MAINMATTER
%%%%%%%%%%%%%%%%%%%%%%%%%%%%%%%%%%%%%%%%%%%%

\section{Introduction}

With the launch of the {\it Fermi} satellite we have entered in a new
era of blazar research. 
The Large Area Telescope (LAT) onboard {\it Fermi} has $\sim$20 fold
better sensitivity than its predecessor EGRET in the 0.1-100 GeV energy range,
enabling the detection of hundreds (and thousands in the end) 
of blazars. 
For the brightest we can study not only their high state,
but also their more normal and quiescent states.
Other key information come from the UVOT, XRT and BAT telescopes onboard
the {\it Swift} satellite, covering the optical and X--ray energy range 
of the blazars detected by {\it Fermi}.
Together with data gathered by ground based telescope, we can routinely 
assemble simultaneous spectral energy distributions (SEDs).
Blazars are extremely variable sources, so having simultaneous snapshot
of the entire SED is a mandatory to start to meaningfully 
explore their physics.
Up to now, there exist two catalogues of {\it Fermi} blazars.
The first \cite{abdo1} lists the $\sim$100 blazars detected 
with high significance ($>10\sigma$) during 
the first three months of all sky survey ({\it Fermi} patrols the entire sky 
every three hours, i.e.
two orbits), while the second, just published \cite{abdo2} list the $\sim$700
blazars detected at more than 4$\sigma$ during the first 11 months of operation.
In addition, there is another catalogue of blazars detected by the {\it Swift}/BAT
instruments in the 15--55 keV energy range \cite{ajello2009}, listing 38 blazars
at high galactic latitudes ($|b<15^\circ|$, see also \cite{cusumano2009} in which
blazars at lower Galactic latitudes are present).

% ------------------------------------------
\begin{figure}
\includegraphics[height=.35\textheight]{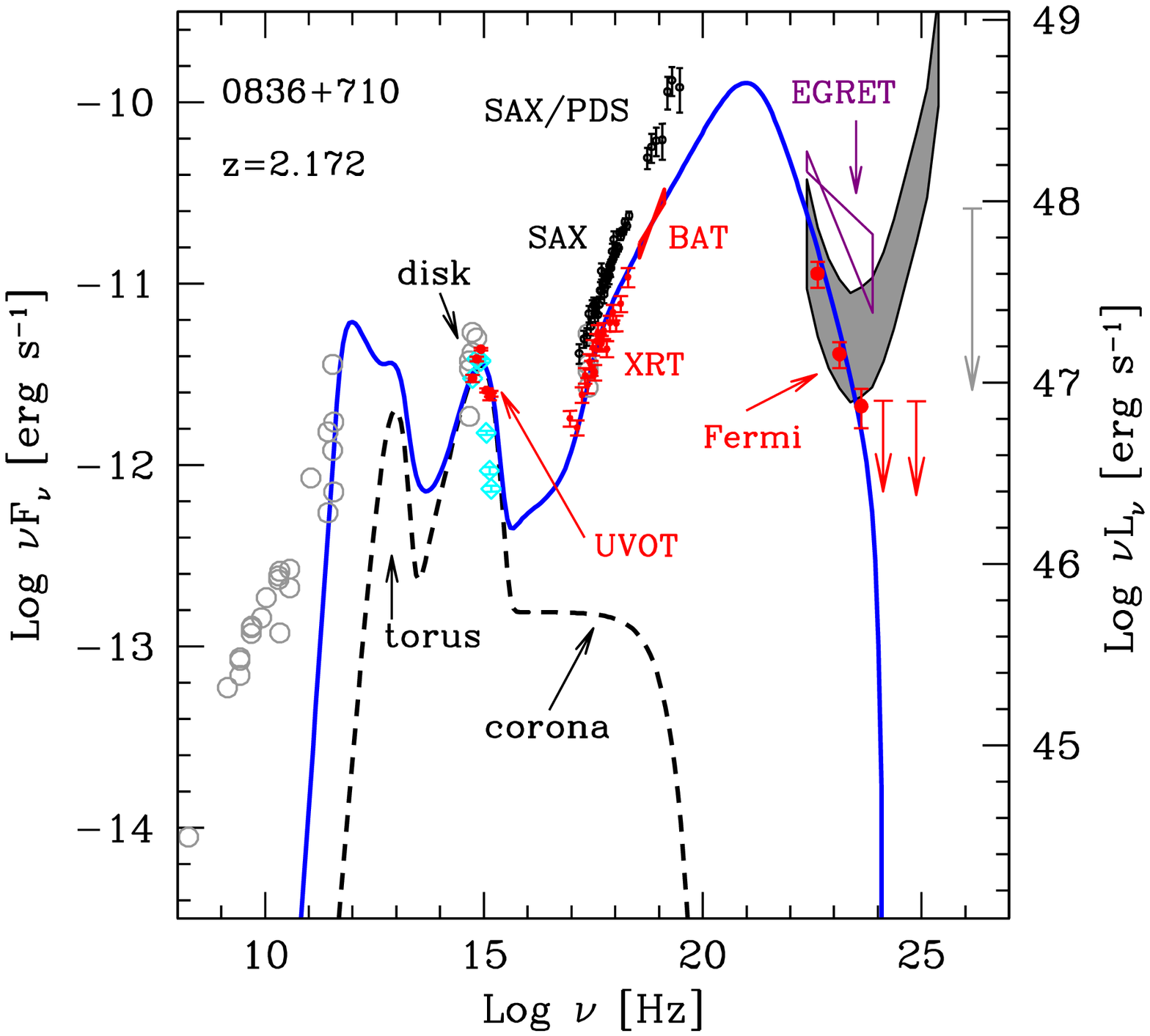}
\includegraphics[height=.35\textheight]{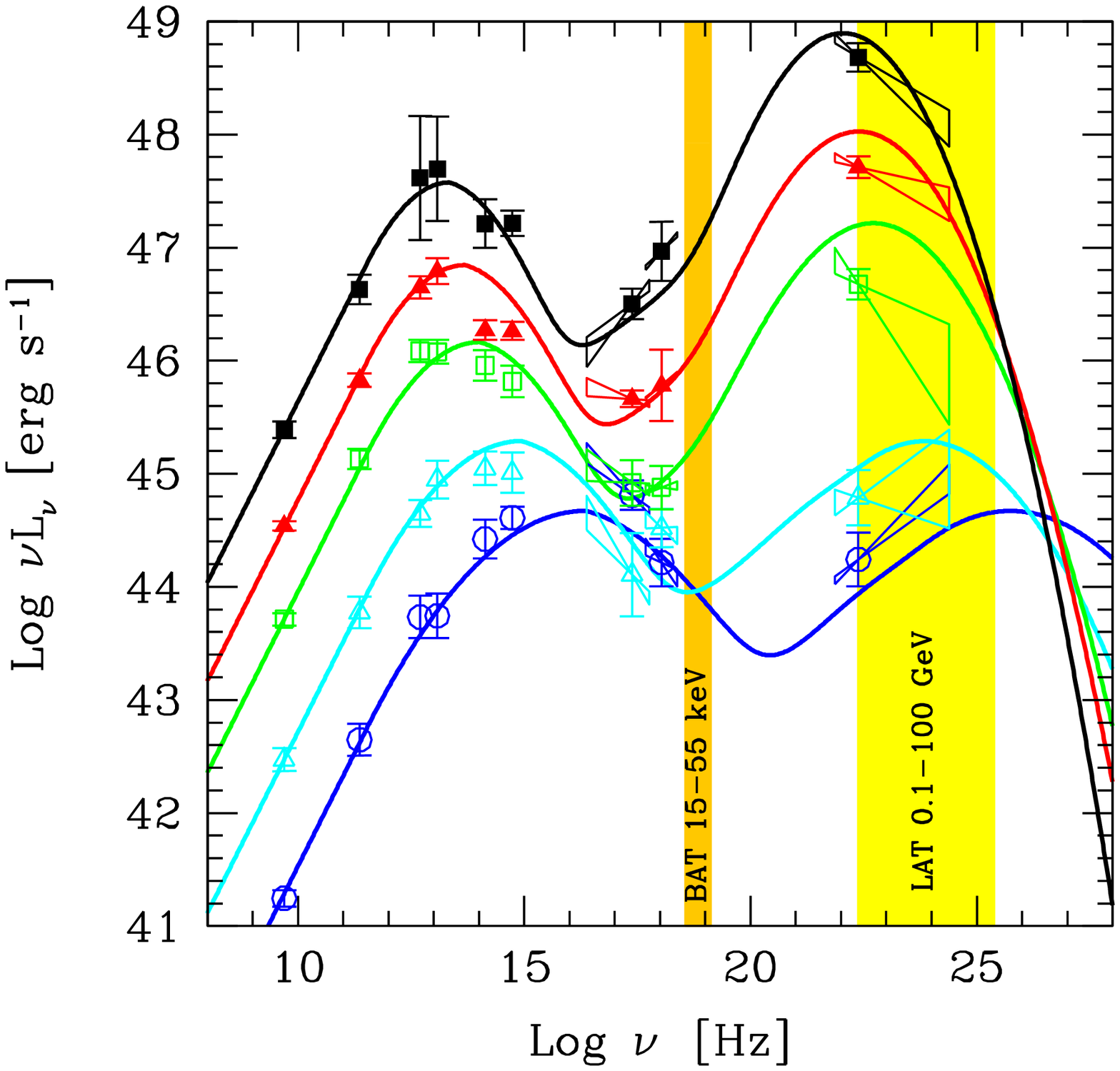}
\caption{Left panel: the SED of the FSRQ 0836+710. 
In this very powerful source the two broad humps produced
by the jet leave the disk component ``naked" and well visible.
The line is a fitting model. Adapted from \cite{chasing}.
Right panel: The blazar sequence 
\cite{fossati1998}, \cite{gg1998}, with overplotted the LAT (0.1--100 GeV) 
and BAT (15--55 keV) energy ranges.
}
\label{sequence}
\end{figure}
% ------------------------------------------

In the following I will report on the results obtained analysing all blazars 
with redshift of the 3--months {\it Fermi} list (85 sources, excluding 4
FSRQs with poor data coverage) and on the 10 FSRQs with $z>2$ 
present in the BAT list.

The novelty, with respect to previous studies, is not only the increased
quantity and quality of the data, but also the realisation that, at least for
high power objects, it is possible to find the black hole mass and the
accretion rate.
To understand how this is possible, consider that all blazars have a SED 
characterised by two broad emission humps,
located at smaller frequencies as the bolometric luminosity increases.
The first hump is thought to be produced by the synchrotron process, while
the high energy one is though to be due to the Inverse Compton process.
At low luminosities, the two humps have more or less the same power,
while the high energy hump becomes more dominant when increasing the 
bolometric luminosity (this is the so called {\it blazar sequence} illustrated 
in Fig. \ref{sequence} \cite{fossati1998}, \cite{gg1998}).
Powerful (broad line emitting) FSRQs have the synchrotron hump
at sub--mm frequencies, and the high energy peak in the MeV band.
They should thus have {\it steep} [$\alpha_\gamma>1$; $F(\nu)\propto \nu^{-\alpha}$]
spectra in the {\it Fermi} band, and {\it flat} spectra ($\alpha_X<1$)
in the BAT band.
If the same electrons are also responsible for the low energy hump, then
the emission above the synchrotron peak should have a spectrum as steep as observed
in the {\it Fermi} band. 
As a consequence, the synchrotron IR--optical--UV flux is weak, and the emission from
the accretion disk can easily dominate in these bands.
The left panel of Fig. \ref{sequence} illustrates this point by showing the 
SED of the powerful FSRQ 0836+710.
By fitting the IR-opt--UV data with a Shakura--Sunjaev \cite{shakura1973} disk
we can find both the black hole mass and the accretion rate. 

At the other extreme of the blazar sequence (i.e. low power BL Lacs),
there is no sign of thermal disk emission, nor of emission lines produced by 
the photo--ionising disk photons.
For them we can only derive an upper limit of the accretion luminosity.

We model the non--thermal SED of all blazars with a one--zone, 
synchrotron and Inverse
Compton leptonic model, to find the physical parameter of the emitting
region of the jet, including the power transported in the form
of particles and fields \cite{ggcanonical}. 

We use a cosmology with $h=\Omega_\Lambda=0.7$ and $\Omega_{\rm M}=0.3$,
and use the notation $Q=Q_X 10^X$ in cgs units (except for the black hole masses,
measured in solar mass units).

\section{The Fermi/LAT and Swift/BAT blazar sequence}

The nearly hundred bright blazars detected by {\it Fermi} in the first 3 months
are sufficiently representative of both the BL Lac and FSRQ populations, and
a good fraction of BL Lacs have a measured redshift.
We can calculate the $\gamma$--ray luminosity for about 90 blazars
(all FSRQs and 75\% of BL Lacs).
We can therefore test a simple consequence of the blazar sequence idea,
namely that FSRQs should have a steeper $\gamma$--ray spectral slope.
This is nicely confirmed \cite{ggdivide}.
The fact there is a ``divide" in $\gamma$--ray luminosity between 
FSRQs and BL Lacs may depend on the narrow range of black hole masses 
and viewing angles explored so far,
since we are detecting the `tip of the iceberg' of a 
broader distribution, extending to smaller black hole masses and slightly
larger viewing angles.
In other words: since our instrumental capability has been limited up to now,
we are bound to detect the most extreme objects in term of $\gamma$--ray luminosities,
that are likely to be associated to sources pointing almost exactly to Earth,
and to the heaviest black hole (this last point will be clearer below).
The idea of the blazar sequence has thus been revisited \cite{gg2008} by including
the effect of a broader distribution of black hole masses, finding, quite simply,
that {\it it is the jet power in units of Eddington that determines the overall shape of
the SED}. 
In absolute terms, therefore, it is not true that low power blazars should have a
``bluer" SED (bluer means having humps peaking at greater energies).
This is so now because we have explored only a relatively small range 
of black hole masses and viewing angles.

\begin{figure}
\includegraphics[height=.35\textheight]{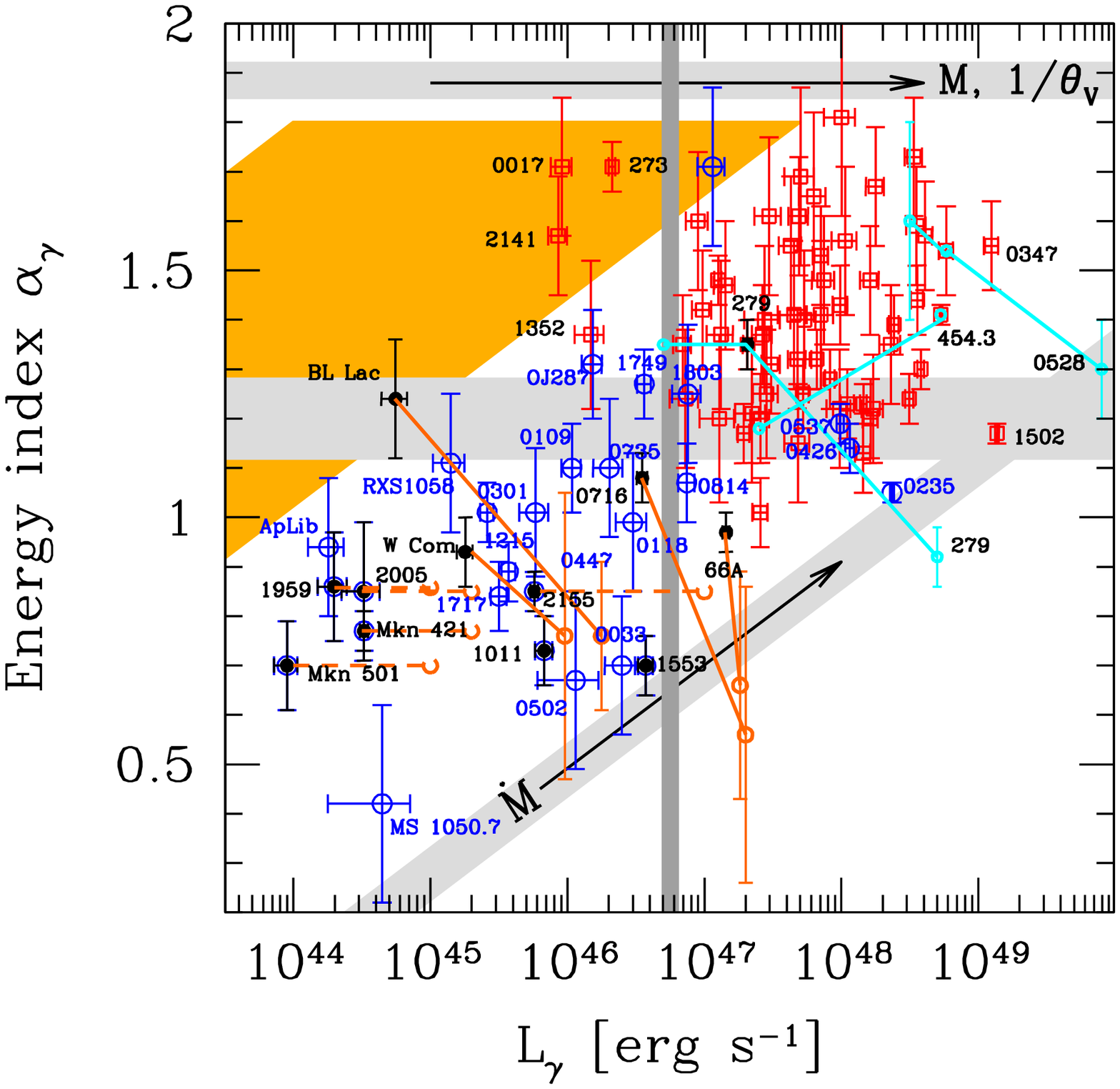}
\includegraphics[height=.35\textheight]{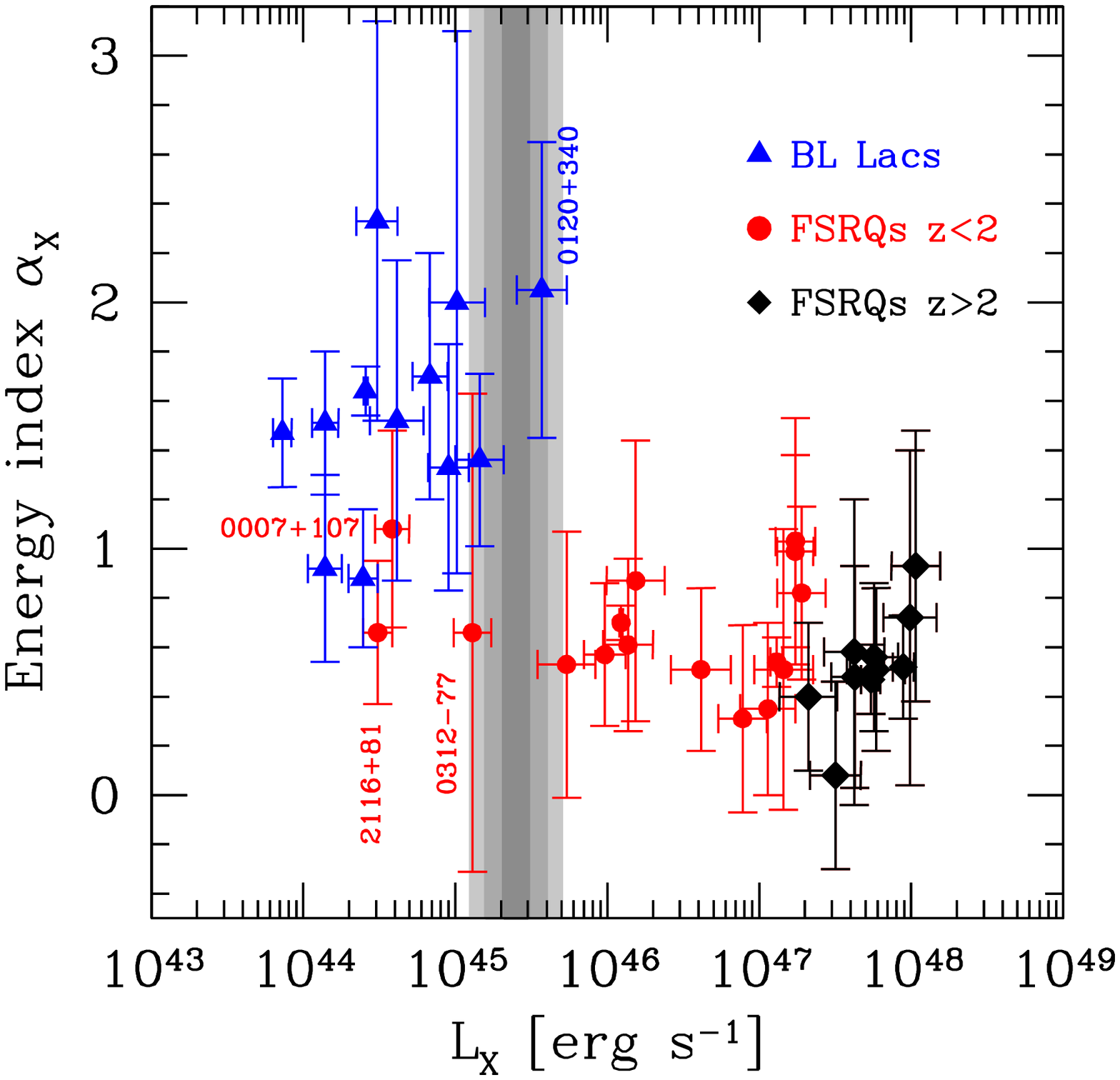}
\caption{{\it Left:} the {\it Fermi} blazars' divide: the $\gamma$--ray 
spectral index as a function of the $\gamma$--ray Luminosity of the blazars
detected by {\it Fermi} during the first 3 months.   
FSRQs are steeper and more luminous than BL Lacs. Adapted from \cite{ggdivide}. 
{\it Right:} the {\it Swift}/BAT blazars' divide: the 15--55 keV spectral
index as a function of the X--ray luminosity for all blazars detected
during the 3 years BAT survey. 
In this plane BL Lacs are {\it steeper} than FSRQs. From \cite{chasing}.
}
\label{divide}
\end{figure}

With this caveat in mind, we can look at the left panel of Fig. \ref{divide}
showing the spectral index of the $\gamma$--ray spectrum as a function of the
$\gamma$--ray luminosity for the bright blazars detected by {\it Fermi} in its 
first 3--months survey.

% ------------------------------------------
% ------------------------------------------

There is a rather well defined boundary between BL Lacs and FSRQs,
and a well defined trend.
This holds {\it despite} the large amplitude variability
of blazars, especially at high energies 
(see how the location of specific sources
can change in Fig. \ref{divide}, as shown by the segments.
Note that several sources ``move" orthogonally to the 
correlation defined by the ensemble of sources, i.e.
they become harder when brighter, with the exception of 3C 454.3).
The high and the low $\gamma$--ray states of single
sources can be dramatically different, and this
implies that the distribution in luminosity 
within each blazar class is largely affected by variability.

Fig. \ref{divide}  shows that BL Lacs and FSRQs separate at
$L_\gamma\sim 10^{47}$ erg s$^{-1}$, as indicated by the vertical 
grey stripe.
Furthermore there is a less clear--cut separation 
in spectral indices, occurring at $\alpha_\gamma\simeq 1.2$
(horizontal grey stripe).

This behaviour is just what the simplest version of the 
``blazar sequence" 
would predict: low power BL Lac objects peak at higher energies,
with the high energy peak often located beyond the LAT range:
they have smaller $L_\gamma$ and flatter $\alpha_\gamma$.
FSRQs, instead, peak at lower frequencies, and the peak of their
high energy emission (dominating their power output) is below 100 MeV.
In the LAT energy range they are steep, but powerful.

The left panel of Fig. \ref{divide} then represents the $\gamma$--ray 
selected version of the blazar sequence, while the right panel is
the hard X--ray version of it, showing a division between {\it steep}
BL Lacs at low power and {\it flat} FSRQs at greater power.
This is because, in low power BL Lacs, the hard X--ray spectrum is often due
to the steep tail of the synchrotron spectrum, while in FSRQs the 
flat part of the inverse Compton component dominates.
Note also that the most distant FSRQs ($z>2$) coincide with the
most X--ray luminous sources.

Going back to the left panel of Fig. \ref{divide}, 
we can wonder what will happen when {\it Fermi} will reach
lower sensitivities as a consequence of the increased exposure time.
Sources of smaller $\gamma$--ray luminosity (at the same redshifts
of the present ones) will become detectable. 
In other words, the top left triangular region of the left
panel of Fig. \ref{divide} will start to be populated.
Will these sources be FSRQs or BL Lacs?
Our expectation is that they will be mostly FSRQs if $\alpha_\gamma>1.2$,
and BL Lacs if $\alpha_\gamma<1.2$. 
{\it These new BL Lacs and FSRQs would on average have a smaller black hole
mass or a slightly larger viewing angle.}

The trend (steeper when brighter) that is now visible {\it will disappear},
since it is due the fact that we now select only the most extreme objects
(smallest viewing angles and largest black hole masses).  
BL Lacs and FSRQs would be mainly divided by their $\alpha_\gamma$ 
(with quite a large region of superposition).
The coloured trapezoidal region in Fig. \ref{divide} will be 
{\it mainly populated by FSRQs 
of smaller black hole mass $M$ or viewing 
angles $\theta_{\rm v}$}:
decreasing $M$ (increasing $\theta_{\rm v}$) we move to the left.
The location of a blazar within each stripe depends instead on the accretion
rate $\dot M$: by increasing it, we increase $L_\gamma$ and steepen $\alpha_\gamma$,
i.e. we move from bottom left to top right.
When doing so, we will cross the ``divide", but we have one dividing line
for each black hole mass (since it corresponds to 
$L_\gamma \propto L_{\rm d}\sim 10^{-2} L_{\rm Edd}$, see below).
Another expectation of the blazar sequence idea is that the region of
flat $\alpha_\gamma$ and high $L_\gamma$ should remain empty, unless 
perhaps for some extraordinary high and transient state of some blazars.

\subsection{The divide}

Fig. \ref{divide} shows a $\gamma$--ray luminosity
dividing BL Lacs from FSRQs.
We suggested \cite{ggdivide} that this is due to a change
of the accretion regime, becoming radiatively inefficient
when the disk emits less than $\sim 10^{-2} L_{\rm Edd}$. 
This is associated to a critical (dividing) luminosity of the 
observed beamed emission, rather well tracked by $L_\gamma$.
To understand why in a simple way, assume
i) that most of the bright blazars detected by the 3--months LAT survey 
have approximately the same black hole mass;
ii) that the largest $L_\gamma$ correspond to jets 
with the largest power carried in bulk motion of particles and fields;
iii) that the jet power and the accretion rate are related.
These three assumptions, that will be better justified below,
imply that the most luminous blazars have the most powerful jets and
are accreting near Eddington. 
These are the FSRQs with $L_\gamma\sim 10^{49}$ erg s$^{-1}$.
Since the dividing $L_\gamma$ is a factor 100 less, it should correspond
to disks emitting 1\% of the Eddington luminosity.
Below this value we find BL Lacs, that have no (or very weak) 
broad emission lines. 
If the disk becomes radiatively inefficient at $L_{\rm d} < 10^{-2}L_{\rm Edd}$
the broad line region receives a much decreased ionising luminosity,
and the lines become much weaker.
The radiation energy density of the lines becomes unimportant for 
the formation of the high energy continuum (there are much less 
seed photons for the Inverse Compton process), implying: 
i) a reduced ``Compton dominance" (i.e. the ratio of the Compton to synchrotron
luminosities);
ii) less severe cooling for the emitting electrons, that can then
achieve larger energies and then
iii) a shift of both the synchrotron and the Inverse Compton peak frequencies 
to larger values.
According to this interpretation, it is the accretion mode that determines
the ``look" of the radiation produced by the jet, not a property of the jet itself.
The very same arguments can be applied to the blazars detected by BAT in 
the hard X--ray band, where the dividing X--ray luminosity is at a few times 
$10^{45}$ erg s$^{-1}$.

We can ask again what will happen when {\it Fermi} will detect less
$\gamma$--ray luminous blazars.
If the idea of a close link between the jet power, the jet luminosity
(as tracked by the $\gamma$--ray one) and accretion is correct, 
these less luminous blazars should have smaller black hole masses.
As a consequence, the dividing line will be located at a smaller $L_\gamma$,
but still corresponding to $L_{\rm d}\sim 10^{-2} L_{\rm Edd}$.

\section{Jets and disks' power}

\begin{figure}
\includegraphics[height=.36\textheight]{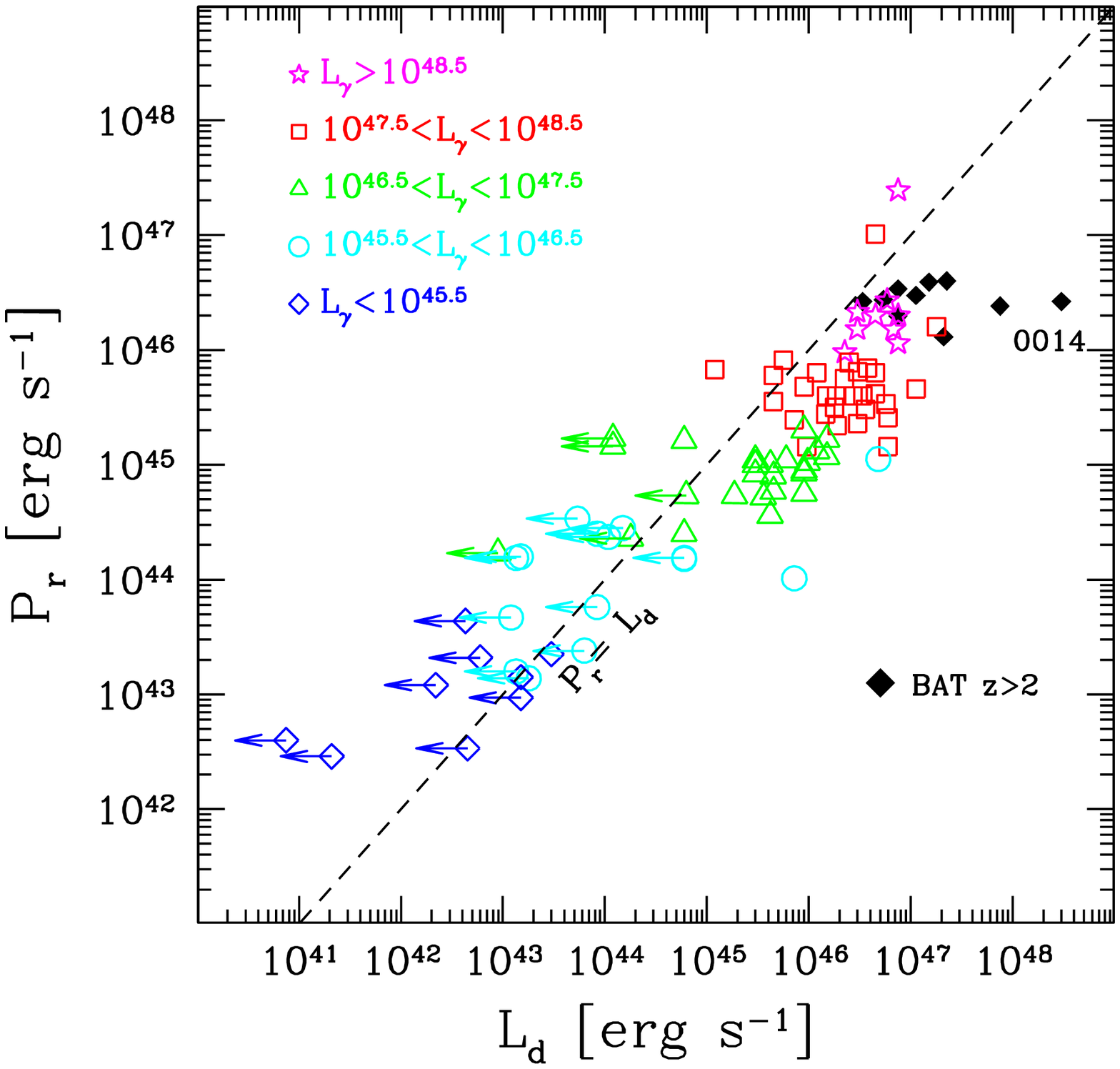}
\includegraphics[height=.36\textheight]{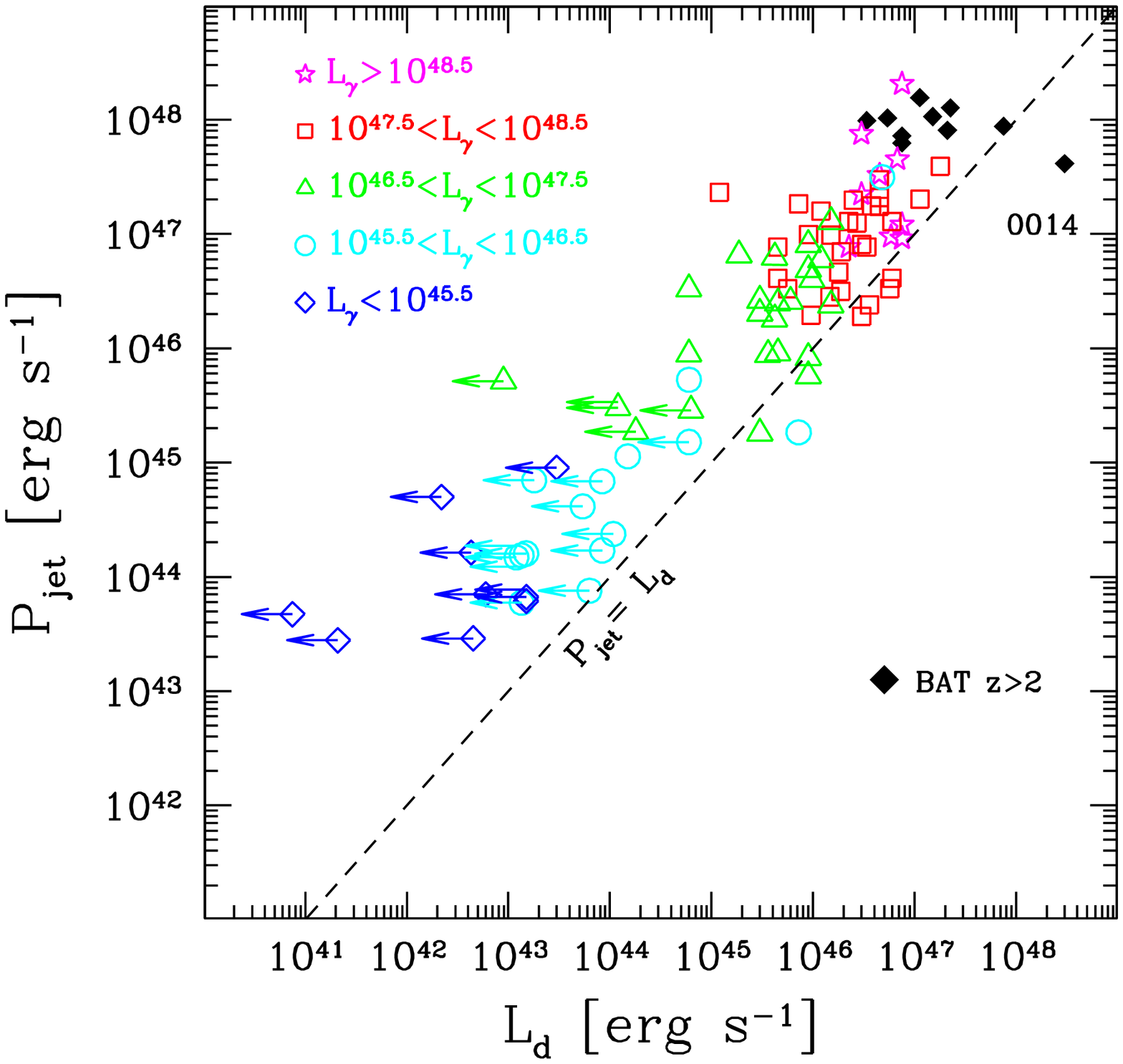}
\caption{
{\it Left:} The power spent by the jet to produce the radiation 
we see, $P_{\rm r}$, as a function
of the accretion disk luminosity $L_{\rm d}$.
BAT blazars (black diamonds) are compared with
the blazars in the {\it Fermi} 3--months catalogue of 
bright sources detected above 100 MeV. 
The latter have different symbols according to their
$\gamma$--luminosity, as labelled.  
{\it Right:} The jet power $P_{\rm jet}$ as a function
of the accretion disk luminosity $L_{\rm d}$.
From \cite{chasing}.
}
\label{power}
\end{figure}

The power of jets can be casts in the form of an energy flux:
\begin{equation}
P_{\rm i} \, =\, \pi R^2 \Gamma^2 c U^\prime_{\rm i}
\end{equation}
where $R$ is the size of the emitting region, $\Gamma$ is the bulk Lorenz factor,
and $U^\prime_{\rm i}$ is the energy density, measured in the comoving frame, 
of radiation [$U^\prime_{\rm r}=L_{\rm syn}/(4\pi R^2 c \delta^4)
+0.1 \Gamma^2 L_{\rm d}/(4\pi R_{\rm BLR}^2 c)]$
emitting electrons [$U^\prime_{\rm e}=m_{\rm e}c^2\int N(\gamma)\gamma d\gamma$], 
magnetic field [$U^\prime_{\rm B}=B^2/(8\pi)$], 
and cold protons [$U^\prime_{\rm p}= 
U^\prime_{\rm e} (m_{\rm e}/m_{\rm p})/\langle \gamma\rangle$, if there is one proton per
emitting electron].
Particularly relevant, and almost model independent, is the power spent
by the jet to produce the radiation we see:
\begin{equation}
P_{\rm r} \, =\, \pi R^2 \Gamma^2 c U^\prime_{\rm r}\, =\, 
L_{\rm obs}\, { \Gamma^2 \over 4  \delta^4} \, \approx \, {L_{\rm obs}\over 4 \Gamma^2}
\end{equation}
where we have assumed $\Gamma\sim\delta$.
$P_{\rm r}$ can be taken as the a strict and robust lower limit to the jet power.
The left panel of Fig. \ref{power} shows $P_{\rm r}$ as a function of the disk 
luminosity $L_{\rm d}$. For BL Lacs, with no signs of disk luminosity, 
$L_{\rm d}$ is an upper limit (arrows in Fig. \ref{power}).
The right panel of Fig. \ref{power} shows the total jet power
$P_{\rm jet}\equiv P_{\rm p}+P_{\rm e}+P_{\rm B}$ as
a function of $L_{\rm d}$.

Considering only FSRQs, there is a correlation between $P_{\rm r}$ and
$L_{\rm d}$, and also between $P_{\rm jet}$ and $L_{\rm d}$
that remain significant also when accounting for the common
redshift dependence. 
The slope of these correlations are consistent with being linear \cite{ggfermi}.
The black diamonds correspond to FSRQs with $z>2$ of the 
{\it Swift}/BAT hard X--ray survey. 
They are the most powerful blazars.

The conclusions we draw from Fig. \ref{power} are that
$P_{\rm r} \sim (1/3) L_{\rm d}$ in FSRQs and larger in BL Lacs,
and that $P_{\rm jet}$ is probably larger than $L_{\rm d}$ for all
blazars, although remaining proportional to $L_{\rm d}$.

% ------------------------------------------
% ------------------------------------------

% ------------------------------------------
\begin{figure}
\includegraphics[height=.36\textheight]{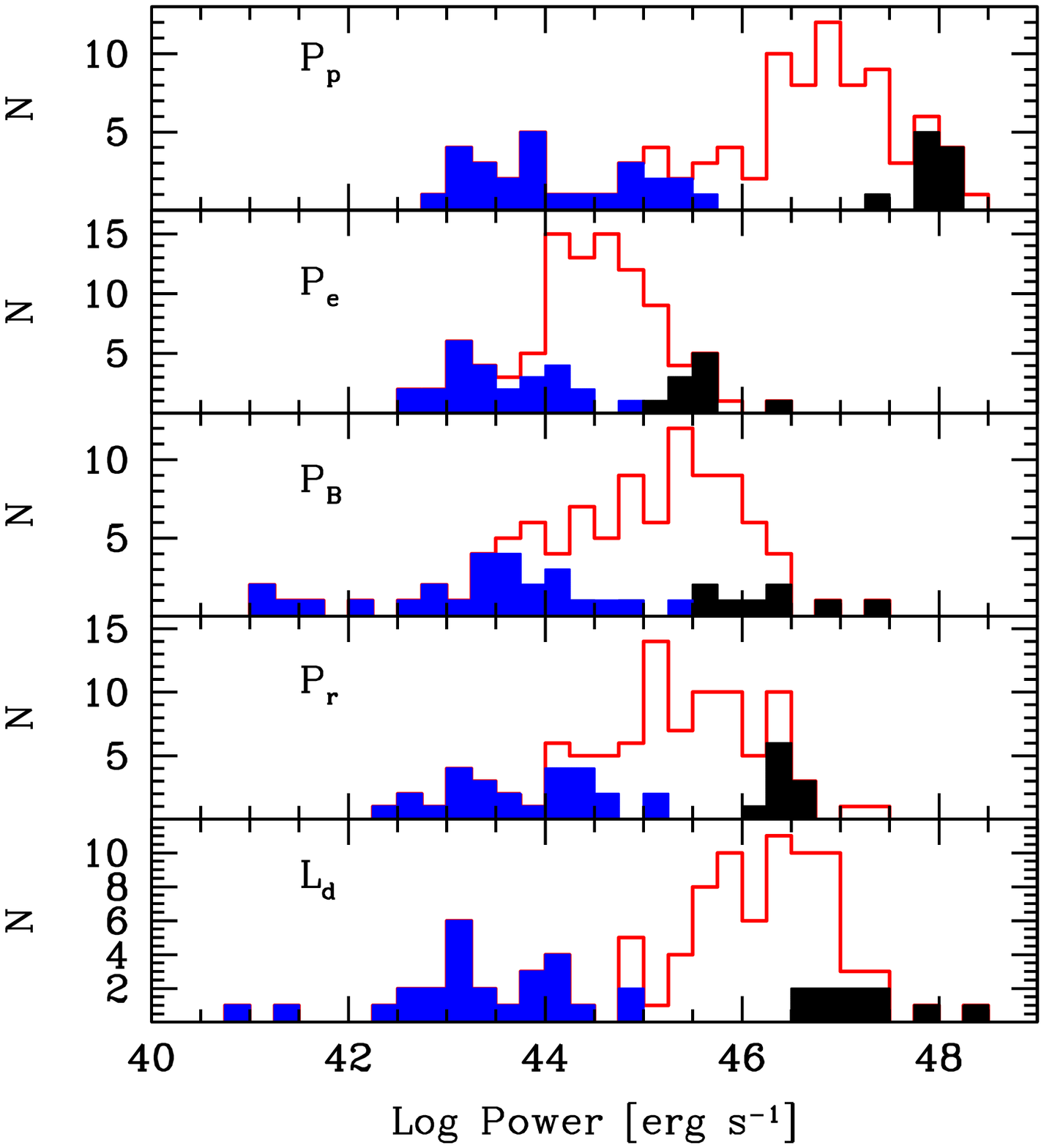}
\includegraphics[height=.36\textheight]{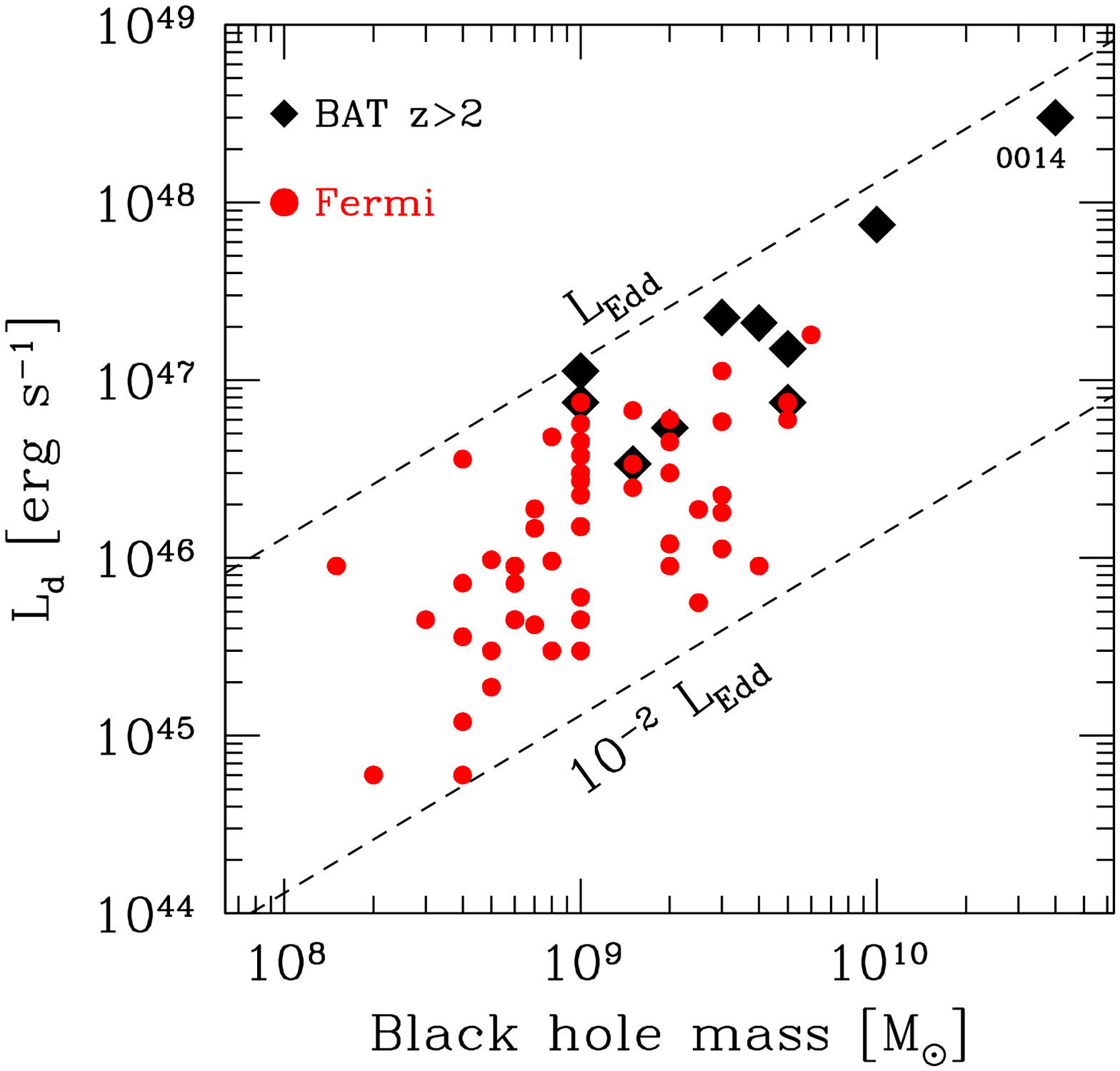}
\caption{
{\it Left:} histograms of the the power carried by the jet and
(bottom panel) of the luminosity of the accretion disk
for all bright {\it Fermi} blazars with redshift and for the 
$z>2$ BAT FSRQs. 
Shaded ares correspond to BL Lacs (left, in blue) and BAT FSRQs (right, in black).
The histogram of $L_{\rm d}$ for BL Lacs corresponds to upper limits.
{\it Right:} the luminosity of the accretion disk, $L_{\rm d}$, as a function of the 
black hole mass for broad emission line FSRQs.  
The dashed lines indicate $L_{\rm d}=L_{\rm Edd}$ and $L_{\rm d}=10^{-2} L_{\rm Edd}$.
The different symbols corresponds to FSRQs in the {\it Fermi}/LAT 3 months survey 
\cite{abdo1} and 
to FSRQs with $z>2$ in the {\it Swift}/BAT 3 years survey \cite{ajello2009}, 
as labelled.
}
\label{isto}
\end{figure}
% ------------------------------------------

\section{Magnetic or matter dominated jets?}

The left panel of Fig. \ref{isto} shows the histograms 
of the power carried by the jet in different forms and 
(at the bottom) the distribution of the luminosity $L_{\rm d}$
produced by the disk (for BL Lacs, the shaded
area corresponds to upper limits on $L_{\rm d}$.)
We have considered all bright {\it Fermi} blazars with redshift
and also (black shaded areas) the 10 FSRQs at $z>2$ detected by
the {\it Swift}/BAT 3 years survey.
With the exception of $P_{\rm r}$ (that, as explained above, depends
only on $1/\Gamma^2$) these jet powers are derived considering how many particles
and magnetic field are needed to account for the radiation we see.
This means that the derived powers are model--dependent: we have adopted 
a leptonic, synchrotron + Inverse Compton model, that has a large consensus
in the scientific community. 
The left panel of Fig. \ref{isto} is the summary of our most recent works on
{\it Fermi}/LAT and {\it Swift}/BAT blazars \cite{ggfermi}, \cite{chasing}.
Previous studies on the jet power not only in blazars
but also in radio--loud objects can be found in 
\cite{rawlings1991}, % Rawlings \& Saunders 1991; 
\cite{celotti1997}, 
\cite{cavaliere2002}, 
\cite{maraschi2003}, % Maraschi \& Tavecchio 2003; 
\cite{sambruna2006}, % Sambruna et al. 2006;
\cite{allen2006}, % Allen et al. 2006;
\cite{acgg2008}, %  Celotti \& Ghisellini 2008;
\cite{gg2008}, % Ghisellini \& Tavecchio 2008; 
\cite{kataoka2008}. % Kataoka et al. 2008).

The power carried in the form of bulk motion of matter depends
on how many emitting electrons we require to account for the SED.  
Since the energy distribution of the emitting electrons is usually
steep, the largest number of electrons is at low energies.
Their synchrotron radiation is invisible, because self--absorbed.
However, for powerful FSRQs, the fast radiative cooling
implies that all particles cool down to low energies
in less than one light crossing time $R/c$.
This is confirmed by fitting the soft X--ray spectra
(that in these sources is due to external Compton, i.e. scatterings
between electrons and broad line photons): the fit requires low
energy electrons. 
Thus for these sources we must assume that the particle
distribution does extend to low energies.

For less powerful BL Lacs the radiative cooling is slower,
and particles may not cool to low energies in the time $R/c$.
However, for these sources, the average electron 
energy needed to explain their SED becomes so large that 
$L_{\rm e} \propto \langle \gamma\rangle$ is only weakly dependent on 
the minimum energy.
Furthermore, for TeV BL Lacs, $\langle \gamma\rangle$ becomes even larger 
than the proton to electron mass ratio, implying that the presence or not 
of cold protons does not change much the total jet power.

Consider first the almost model--independent power $P_{\rm r}$, namely 
the power spent by the jet to produce its radiation. 
For FSRQs, $P_{\rm r}$ extends to larger values than the distribution 
of $P_{\rm e}$, the power carried by the jet in the form
of emitting electrons.
So the jet cannot be made by electrons and pairs only.
Can the needed power come from the Poynting flux (by e.g. reconnection)?
The distribution of $P_{\rm B}$ is at slightly smaller values than the 
distribution of $P_{\rm r}$, indicating that the Poynting flux cannot
be at the origin of the radiation we see.
As described in \cite{acgg2008}, this is a direct consequence
of the large values of the Compton dominance (i.e. the ratio of 
the Compton to the synchrotron luminosity is small), since
this limits the value of the magnetic field.

To justify the power that the jet carries in radiation we are forced 
to consider protons.
If there is one proton per electron (i.e. no pairs), then
$P_{\rm p}$ for FSRQs is a factor $\sim$10--100 larger than $P_{\rm r}$,
meaning an efficiency of 1--10\% for the jet to convert its bulk kinetic
motion into radiation (see also \cite{sikora2000} about this point).
This is reasonable: most of the jet power in FSRQs goes to form and energize the 
large radio structures, and not into radiation.
                       
We then conclude that jets should be matter dominated, at least at the 
scale where most of their luminosity is produced 
(about $10^3$ Schwarzschild radii from the black hole).
At the jet start, instead, it is likely that the magnetic field is dominant
and responsible for the acceleration of the jet.
Therefore the acceleration
of jet, if magnetic, must be very efficient, corresponding to sub--equipartition
magnetic fields when the jet reaches its final velocity, at distances
smaller than $\sim 10^3$ Schwarzschild radii.

\section{Black hole masses}

When the accretion disk luminosity is visible (i.e. in FSRQs),
we directly derive the black hole mass.
Uncertainties in the mass value depends on the quality of data, 
on the assumption of isotropic emission and on the assumption
of a standard Shakura--Sunjaev disk with a total disk luminosity 
$L_{\rm d} =\eta \dot M c^2$, with $\eta=0.08$.
The assumption of isotropic emission is questionable at very large and
very low accretion rates, in Eddington units, for which the disk
can become geometrically thick, hotter, less radiatively
efficient (i.e. smaller $\eta$) and develop a funnel in its
internal part. The funnel can collimate the radiation
along the axis, therefore, for blazars, along the line of sight.
For very small accretion rates these effects have no influence
on our estimates: we derive only upper limits on $\dot M$ 
and the values of the masses are very uncertain in any case.
For $L_{\rm d}$ approaching $L_{\rm Edd}$, instead, we may 
overestimate the disk luminosity.
Indeed, in Fig. \ref{power} we have labelled the location of S5 0014+813,
that seems to be an outlier with respect to the $P_{\rm r}$--$L_{\rm d}$ and
$P_{\rm jet}$--$L_{\rm d}$ relations defined by the other FSRQs.
It is also the object with the heaviest black hole: 40 billion solar masses.
If the disk luminosity of this source is somewhat collimated by a funnel,
and therefore the flux enhanced in our direction,
then we have overestimated its $M$ and $\dot M$.
If its true disk luminosity were a factor $\sim$10 less, then this 
source would join the jet/disk power relations.

On the other hand, the fact that the other FSRQs define a 
correlation, although they have different $L_{\rm d}/L_{\rm Edd}$,
argues against strong collimation effects causing a 
scatter larger than observed.

The right panel of Fig. \ref{isto} shows the disk luminosity
as a function of the derived black hole mass for the {\it Fermi}/LAT
and for the $z>2$ {\it Swift}/BAT FSRQs.
One can see that the {\it Swift}/BAT FSRQs have more luminous disk, and 
larger black hole masses.
After accounting for the different sensitivities of LAT and BAT, we arrive
to conclude that 
{\it hard X--ray surveys are the most efficient way to find the most extreme blazars
in terms of black hole mass, disk luminosity and jet power.}
This fact will be very important for the new planned hard X--ray missions,
such as {\it NHXM}\footnote{http://www.brera.inaf.it/NHXM2/} and 
{\it EXIST}\footnote{http://exist.gsfc.nasa.gov/}.

\begin{figure}
\includegraphics[height=.35\textheight]{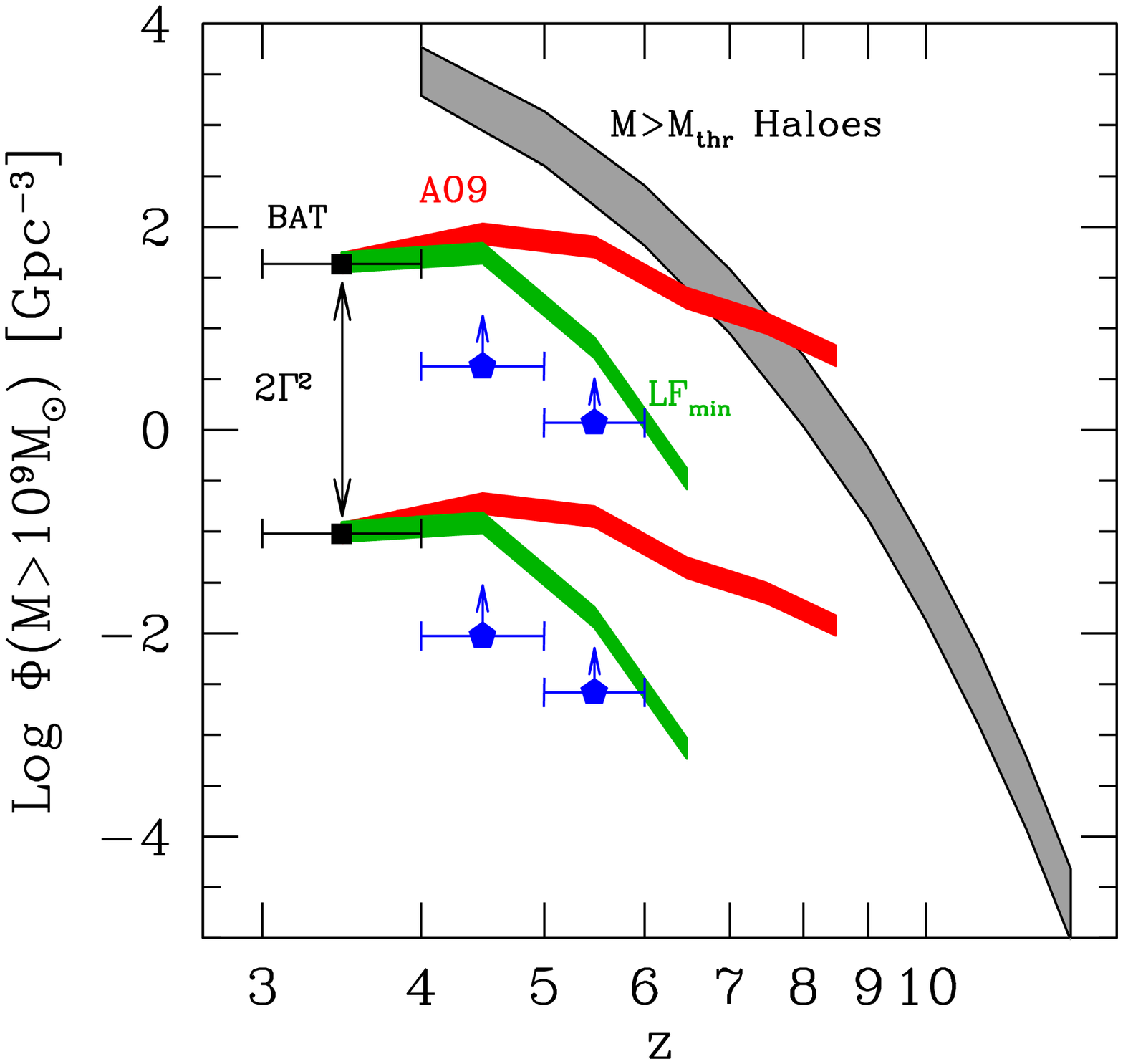}
\includegraphics[height=.33\textheight]{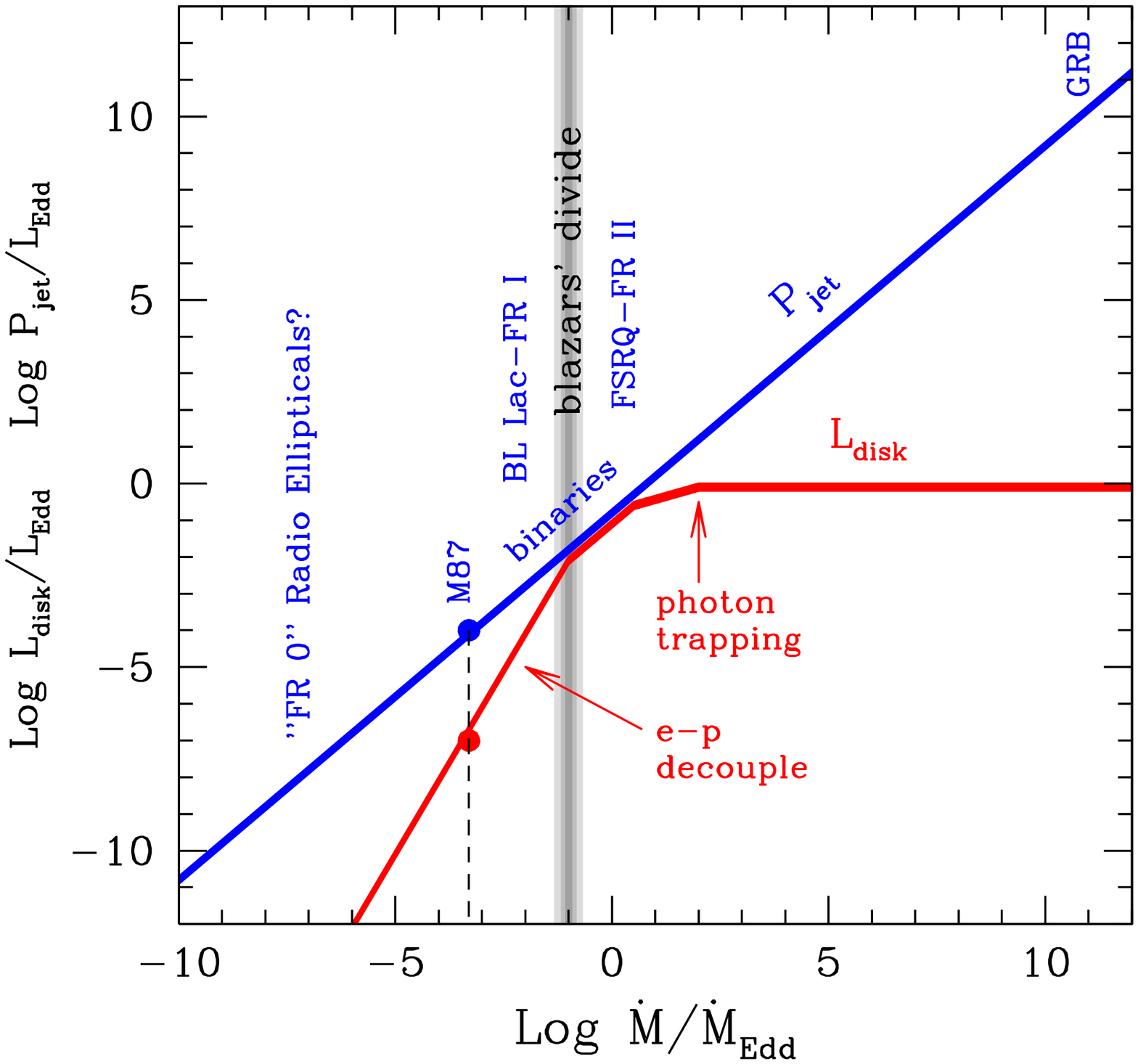}
\caption{
{\it Left:} Number density of black holes with $M>10^9 M_\odot$ as a function of redshift. 
The grey stripe is based on associating black hole mass to halo mass.
In the lower part of the figure we report, as grey stripes (red and green in the
electronic version), the mass function
$\Phi(z,M>10^9M_\odot)$ for blazars as derived from the luminosity function of 
\cite{ajello2009}. 
The upper (red) stripe corresponds to the 
cosmological evolution model of \cite{ajello2009} extrapolated up to $z \sim 9$, 
while the lower stripe labelled ``minimal LF" corresponds to a less extreme evolution.
The filled pentagons and arrows are the lower limits derived from
the existence of a few blazars in the 2 redshift bins.
In the upper part of the figure we show the same points/stripes
up--shifted by the factor $2\Gamma^2=450$, to account for misaligned sources.
In this way the upper stripe of the mass function
$\Phi(z,M>10^9M_\odot)$ for radio--loud sources is in conflict,
at high $z$, with the estimates derived by massive halos, while
the lower (green) stripe derived through the ``minimal LF" is consistent.
From \cite{chasing}.
{\it Right:} schematic view of how the disk luminosity and the jet power are
related to the accretion rate (all quantities in Eddington units. Beware that 
$\dot M_{\rm Edd}\equiv L_{\rm Edd}/c^2$, i.e. without the efficiency term).
In this plane we can locate all the jetted sources, 
including the radio--galaxies, and the jetted X--ray binaries,
that can cross the `divide', having very different emission states.
Note the location of M87, with a very weak jet and very sub--Eddington disk,
and the location of the ``FR 0" radio--galaxies, having 
an extended emission $\sim$100 times weaker than in FR I radio--galaxies
of similar radio--cores luminosities \cite{baldi2009}.
}
\label{mass}
\end{figure}

All FSRQs with $L_X>2\times 10^{47}$ erg s$^{-1}$ in the 15--55 keV
band have $z>2$ and have black hole masses exceeding $10^9 M_\odot$. 
This allows to use the high luminosity end of the  
hard X--ray luminosity function derived
by \cite{ajello2009} to find the blazar mass function $\Phi(z, M>10^9M_\odot)$,
namely the density of blazar black holes heavier than one billion of
solar masses as a function of redshift.
This is shown in the left panel of Fig. \ref{mass}.

% ------------------------------------------
% ------------------------------------------

This figure also shows, as a grey stripe, the black hole mass
function derived through the density of dark haloes 
(see \cite{chasing} for details) assuming that the relation between 
the mass of the halo and the mass of 
the associated black hole is the same as found at low redshifts.
Since this overestimates the mass of the black holes (at high redshifts they are still 
growing) we can assume that the grey stripe is a sort of upper limit to
$\Phi(z, M>10^9M_\odot)$.
The lower (red) stripe is the $\Phi(z, M>10^9M_\odot)$ derived from
the luminosity function of \cite{ajello2009}.
The lower (green) stripe is the $\Phi(z, M>10^9M_\odot)$
derived modifying the luminosity function of \cite{ajello2009}
by assuming less evolution above $z=4$, where we have no data.
We have called this ``minimal luminosity function", since it is consistent
with all existing data, but minimises the number of heavy black holes
(and powerful X--ray blazars) at high redshifts.
The blue arrows corresponds to the existence of powerful X--ray blazars 
(with large black hole masses) in the two redshift bins.
They define a lower limit because they have not been 
discovered through an all sky hard X--ray survey, 
so others blazars may exist.

The upper red and green stripes take into account that, for each blazar
pointing at us, there are other $\sim 2\Gamma^2=450(\Gamma/15)^2$ blazars
pointing in other directions. 
Note that the $\Phi(z, M>10^9M_\odot)$ derived from \cite{ajello2009}
crosses the ``upper limit" defined by the dark--halo argument,
while the green stripe does not.

To summarise: the BAT blazar survey allowed to meaningfully
construct the hard X--ray LF of blazars.
Its high luminosity end can be translated into the mass function 
of black holes with more than one billion solar masses.
Up to $z=4$, where we do see blazars, the cosmological evolution model, 
as derived by \cite{ajello2009}, is secure.
Beyond $z=4$ it depends strongly on the assumed evolution.
We have then constructed the minimal evolution
consistent with the existing data and the (few)
existing lower limits.
As Fig. \ref{mass} shows, the true mass function of heavy black
hole in jetted sources should be bracketed by the two shown
mass functions derived from \cite{ajello2009} and the ``minimal" LF.
The true mass density should then lie in--between the two
possible choices.

\section{Final considerations}

The right panel of Fig. \ref{mass} is a sketch trying to summarise
how the jet power and the disk luminosity behaves {\it for all
sources having a relativistic jet}, namely blazars and radio--galaxies,
jetted galactic X--ray binaries and Gamma Ray Bursts.
It shows $L_{\rm d}$ and $P_{\rm jet}$ in Eddington units
as a function of the accretion rate, again in Eddington units,
where we have defined $\dot M_{\rm Edd}\equiv L_{\rm Edd}/c^2$,
i.e. without the efficiency $\eta$ (that is changing).
Therefore, at the blazar divide, $L_{\rm d}/L_{\rm Edd}=10^{-2}$
corresponds to $\dot M/\dot M_{\rm Edd}=10^{-1}$.
The scenario we are proposing is that $P_{\rm jet}$ is always
of the order of $\dot M c^2$. 
Instead $L_{\rm d} \propto \dot M^2$
below $\dot M\sim 0.1 \dot M_{\rm Edd}$ (corresponding to the blazar
divide $L_{\rm d}=10^{-2} L_{\rm Edd}$) 
\cite{narayan1997}, when the low densities of the accreting matter 
makes protons and electrons to decouple; becomes the usual
$L_{\rm d}\sim 0.1 \dot M c^2$ at intermediate values, and
almost constant when the accretion rate becomes larger than
$\sim 10\, \dot M_{\rm Edd}$, when the photons produced
inside the disk remain trapped and are swallowed by the hole before
escaping.
We have also included, in Fig. \ref{mass}, a relatively new class
of radio--galaxies, having much less extended radio luminosities than
FR Is of the same radio core luminosities \cite{baldi2009}.
We call these sources ``FR 0" radio--galaxies.
In these scheme, they should correspond to jet powers even smaller than
in FR I sources. These jets can be decelerated easily, before
reaching larger scales.
These sources should have accretion disk accreting at very sub--Eddington rates.
Therefore ``FR 0" and FR I radiogalaxies and their aligned counterpart
(BL Lacs) should have jets much more powerful than the luminosities of their 
accretion disks.

\noindent
{\bf An example: M87} The nearby FR I radio--galaxy, with its visible and
$\gamma$--ray emitting jet, but with a rather strong limit to its accretion
luminosity, can illustrate our point. 
It has a black hole of $6\times 10^9 M_\odot$, a disk luminosity
$L_{\rm d}< 10^{41}$ erg s$^{-1}$ and a jet power  
$P_{\rm jet}\sim 2\times 10^{44}$ erg s$^{-1}$ \cite{tavecchiom87}.
Therefore $L_{\rm d}/L_{\rm Edd} < 10^{-7}$ and 
$P_{\rm jet}/L_{\rm Edd} \sim 2.5\times 10^{-4}$.
These values are plotted in the right panel of Fig. \ref{mass},
and are just on top of the plotted lines.

In summary, our main conclusions are the following:

\begin{itemize}

\item Blazar jets are powerful. The power they transport
to large distances can be even greater then the luminosity produced by the 
associated accretion disk.

\item The jet power is proportional to the accretion rate $\dot M$.
Therefore there is a very close link between jets and disks.

\item The division between line emitting FSRQs and weak line BL Lacs 
reflects the change of the regime of accretion, occurring at roughly
$L_{\rm d}\sim 10^{-2} L_{\rm Edd}$.
Above this value the disk can efficiently photo--ionise the 
Broad Line region, below this value the disk becomes radiatively 
inefficient and the broad lines becomes very weak or absent.

\item Despite this change in the accretion properties, the 
jet power remains of the order of $P_{\rm jet} \sim \dot M c^2$.
Jets track $\dot M$ better than disks.

\item If $\dot M$ evolves in time, also FSRQs and BL Lacs do.
FSRQs evolves positively, BL Lacs negatively. 
FSRQs (FR II) should transform into BL Lacs (FR I) 
if $\dot M$ decreases in time.

\item Blazar jets are matter, not magnetically, dominated.
Since it is conceivable that they are accelerated by magnetic
forces at the start, this implies that the acceleration 
mechanism is very efficient (i.e. the initial Poynting flux
is quickly converted in bulk motion of matter).

\item Bulk motion of cold protons, not electron--positron pairs,
is the dominant component of the jet power.
 
\item The most efficient way to find large black hole masses in radio--loud AGNs
at high redshifts is through hard X--ray surveys.
In fact, increasing the jet power, the hard X-ray luminosity 
increases more than the $\gamma$--ray one. 
Future planned satellites such that the {\it New Hard X-ray Mission, NHXM} and
{\it EXIST} can then find the heaviest black holes at the largest
redshifts.

\end{itemize}

%%%%%%%%%%%%%%%%%%%%%%%%%%%%%%%%%%%%%%%%%%%%%%%%
%% BACKMATTER
%%%%%%%%%%%%%%%%%%%%%%%%%%%%%%%%%%%%%%%%%%%%%%%%

\begin{theacknowledgments}
I gratefully thank my collaborators A. Celotti, R. Della Ceca, 
L. Foschini, G. Ghirlanda, F. Haardt, L. Maraschi, G. Tagliaferri, 
F. Tavecchio and M. Volonteri.
I thank the grant PRIN--INAF 2007 for partial funding.
\end{theacknowledgments}



\bibliographystyle{aipproc}    % if natbib is available
% \bibliographystyle{aipprocl} % if natbib is missing

\bibliography{sample}

%  \typeout{******************************************}
%  \typeout{** Please run "bibtex \jobname" to optain}
%  \typeout{** the bibliography and then re-run LaTeX}
%  \typeout{** twice to fix the references!}
%  \typeout{******************************************}

% \end{document}

%%%%%%%%%%%%%%%%%%%%%%%%%%%%%%%%%%%%%%%%%%%
%% The following lines show an example how to produce a bibliography
%% without the help of the BibTeX program. This could be used instead
%% of the above.
%%%%%%%%%%%%%%%%%%%%%%%%%%%%%%%%%%%%%%%%%%%

% \endinput
\end{document}